\documentclass[prb,twocolumn,showpacs,preprintnumbers,amsmath,amssymb,floatfix]{revtex4}

\usepackage{graphicx,color,bm}

\begin{document}

\title{Electrically tunable band gap in silicene}

\author{N.\ D.\ Drummond}

\affiliation{Department of Physics, Lancaster University, Lancaster LA1 4YB,
  United Kingdom}

\author{V.\ Z\'{o}lyomi}

\affiliation{Department of Physics, Lancaster University, Lancaster LA1 4YB,
  United Kingdom}

\author{V.\ I.\ Fal'ko}

\affiliation{Department of Physics, Lancaster University, Lancaster LA1 4YB,
  United Kingdom}

\date{\today}

\begin{abstract} We report calculations of the electronic structure of
  silicene and the stability of its weakly buckled honeycomb lattice in an
  external electric field oriented perpendicular to the monolayer of Si
  atoms. The electric field produces a tunable band gap in the Dirac-type
  electronic spectrum, the gap being suppressed by a factor of about eight by
  the high polarizability of the system.  At low electric fields, the
  interplay between this tunable band gap, which is specific to electrons on a
  honeycomb lattice, and the Kane-Mele spin-orbit coupling induces a
  transition from a topological to a band insulator, whereas at much higher
  electric fields silicene becomes a semimetal.  \end{abstract}


\maketitle


\section{Introduction \label{sec:intro}}

Two-dimensional (2D) carbon crystals are hosts for Dirac-type electrons, whose
unusual properties have been studied extensively in graphene monolayers
produced by mechanical exfoliation from
graphite.\cite{novoselov_2004,geim_2007} A close relative of graphene, a 2D
honeycomb lattice of Si atoms called \textit{silicene},\cite{cahangirov} does
not occur in nature, but nanoribbons of silicene have been synthesized on
metal surfaces.\cite{exp01,exp02,exp03} Due to the similarity of the lattice
structures, the band structure of silicene resembles that of graphene,
featuring Dirac-type electron dispersion in the vicinity of the corners of its
hexagonal Brillouin zone (BZ)\@.\cite{wallace} Moreover, silicene has been
shown theoretically to be metastable as a free-standing 2D
crystal,\cite{cahangirov} implying that it is possible to transfer silicene
onto an insulating substrate and gate it electrically.  In this work we
predict the properties of this 2D crystal.

\begin{figure}
\begin{center}
\includegraphics[width=0.45\textwidth]{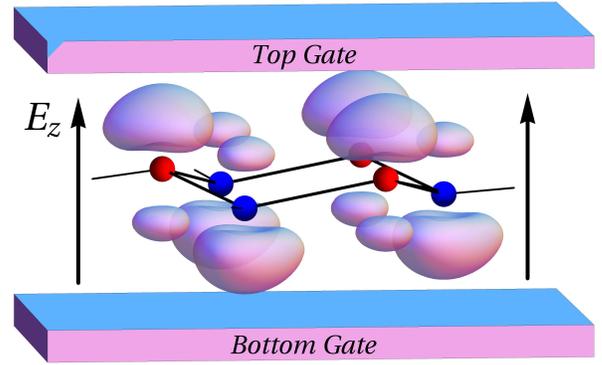}
\caption{(Color online) Atomic structure of silicene, together with a sketch
  of the charge density for the highest occupied valence band in the vicinity
  of the K point.
\label{fig:silicene_in_capacitor}}
\end{center}
\end{figure}

The similarity between graphene and silicene arises from the fact that C and
Si belong to the same group in the periodic table of elements.  However, Si
has a larger ionic radius, which promotes sp$^3$ hybridization, whereas sp$^2$
hybridization is energetically more favorable in C\@.  As a result, in a 2D
layer of Si atoms, the bonding is formed by mixed sp$^2$ and sp$^3$
hybridization.  Hence silicene is slightly buckled, with one of the two
sublattices of the honeycomb lattice being displaced vertically with respect
to the other, as shown in Fig.\ \ref{fig:silicene_in_capacitor}.  Such
buckling creates new possibilities for manipulating the dispersion of
electrons in silicene and opening an electrically controlled
sublattice-asymmetry band gap.\cite{ni2012} In this article we report density
functional theory (DFT) calculations of the band gap $\Delta$ for Dirac-type
electrons in silicene opened by a perpendicular electric field using a
combination of top and bottom gates. We show that $\Delta$ can reach tens of
meV before the 2D crystal transforms into a semimetal and then, at still
higher fields, loses structural stability.  We also determine the weak
electric field at which electrons in silicene experience a transition from a
topological insulator regime\cite{hasan_kane,qi_2011} caused by the Kane-Mele
spin-orbit (SO) coupling\cite{kane_mele} for electrons on a honeycomb lattice
into a conventional band insulator regime.

The rest of this article is arranged as follows. In
Sec.\ \ref{sec:struct_electronic} we report our results for the structural and
electronic properties of free-standing silicene, and compare them with other
theoretical and experimental results in the literature.  In
Sec.\ \ref{sec:transverse_field} we analyze the effects of a transverse
electric field on the structural and electronic properties of silicene, and in
Sec.\ \ref{sec:spin_orbit} we discuss the effects of SO coupling on the
electronic structure, arguing that a crossover from topological insulating
behavior to band insulating behavior must take place as the transverse field
increases in strength.  In Sec.\ \ref{sec:comp_meth} we give the technical
details of our computational methodology and demonstrate the convergence of
our results with respect to simulation parameters.  Finally, we draw our
conclusions in Sec.\ \ref{sec:conclusions}.

\section{Structural and electronic parameters of free-standing
  silicene \label{sec:struct_electronic}}

\subsection{Comparison with theoretical and experimental results in the
  literature}

The lattice constant and the $z$ (out-of-plane) coordinates of the Si atoms
lying on the 2D honeycomb lattice were both fully relaxed using DFT (i) in the
local density approximation (LDA), (ii) with the Perdew-Burke-Ernzerhof (PBE)
exchange-correlation functional,\cite{pbe} and (iii) with the screened
Heyd-Scuseria-Ernzerhof 06 (HSE06) hybrid functional.\cite{hse,hse2} Our DFT
calculations were performed using the \textsc{castep}\cite{castep,castep_dfpt}
and \textsc{vasp}\cite{vasp} plane-wave-basis codes, using ultrasoft
pseudopotentials and the projector-augmented-wave (PAW) method, respectively.
The $z$-coordinates of the two Si atoms in the unit cell (the $A$ and $B$
sublattices) differ by a finite distance $\Delta z$. Our results for a free
silicene monolayer are shown in Table \ref{table:parameters_summary}.  The
metastable lattice that we find is the same as the ``low-buckled'' structure
found by Cahangirov \textit{et al.}\cite{cahangirov}

\begin{table}
\begin{center}
\caption{Silicene structural and electronic parameters: lattice constant $a$,
  sublattice buckling $\Delta z$ (the difference between the $z$ coordinates
  of the $A$ and $B$ sublattices), cohesive energy $E_c$, and Fermi velocity
  $v$. The calculated cohesive energy of silicene includes the DFT-PBE
  zero-point energy, which we found to be $0.10$ eV per atom.  The theoretical
  results are for free-standing silicene; the experimental results are for
  silicene nanoribbons on Ag substrates.
  \label{table:parameters_summary}}
\begin{tabular}{lcccc}
\hline \hline

Method & $a$ ({\AA}) & $\Delta z$ ({\AA}) & $E_c$ (eV) & $v$ ($10^5$
ms$^{-1}$) \\

\hline

PBE (\textsc{castep}) & $3.86$ & $0.45$ & $4.69$ & $5.27$ \\

PBE (\textsc{vasp})   & $3.87$ & $0.45$ & $4.57$ & $5.31$ \\

PBE\cite{ni2012}      & $3.87$ & $0.46$ &        &   \\

LDA (\textsc{castep}) & $3.82$ & $0.44$ & $5.12$ & $5.34$ \\

LDA (\textsc{vasp})   & $3.83$ & $0.44$ & $5.00$ & $5.38$ \\

LDA\cite{cahangirov} & $3.83$ & $0.44$ & $5.06$ & $\approx 10$ \\

LDA\cite{pan2011}     & $3.86$ & $0.44$ &        &   \\

HSE06 (\textsc{vasp}) & $3.85$ & $0.36$ & $4.70$ & $6.75$ \\

Exp.\ [on Ag(110)]\cite{exp02} & $3.88$ & & \\

Exp.\ [on Ag(111)]\cite{exp03} & $3.3$~ & $0.2$~ & \\

\hline \hline
\end{tabular}
\end{center}
\end{table}

The experimental results for the lattice parameter depend on the choice of
substrate on which the silicene is grown.\cite{exp02,exp03} The extent to
which theoretical results obtained for free-standing silicene are applicable
to the silicene samples that have been produced to date is therefore unclear.

\subsection{Stability of free-standing silicene \label{sec:phonon_results}}

\begin{figure}
\begin{center}
\includegraphics[clip,width=0.45\textwidth]{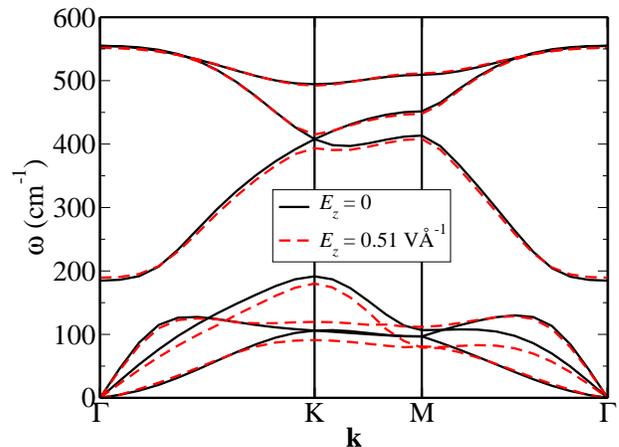}
\caption{(Color online) DFT-PBE phonon dispersion curves for silicene in zero
  external field and at $E_z=0.51$ V{\AA}$^{-1}$.  In both cases the
  calculations were performed using the method of finite displacements, with
  the atomic displacements being 0.0423 {\AA}, in a supercell consisting of
  $3\times 3$ primitive cells with a $20 \times 20$ ${\bf k}$-point grid in
  the primitive cell.
\label{fig:phonons_in_field}}
\end{center}
\end{figure}

The cohesive energy of bulk Si (including a correction for the zero-point
energy) has been calculated within DFT-LDA as 5.34 eV\@.\cite{alfe_2004}
Comparing this with our DFT-LDA cohesive energy of silicene reported in Table
\ref{table:parameters_summary} shows that bulk Si is substantially (0.22 eV
per atom) more stable than silicene, implying that silicene would not grow
naturally as a layered bulk crystal like graphite.  However, by calculating
the DFT phonon dispersion it has been verified both here and in
Ref.\ \onlinecite{cahangirov} that the structure is dynamically stable: no
imaginary frequencies appear anywhere in the BZ\@.  The results of such an
analysis are summarized in Fig.\ \ref{fig:phonons_in_field}. This convinces us
that, as a metastable 2D crystal, silicene can be transferred onto an
insulating substrate, where its electronic properties can be studied and
manipulated as suggested below.

\subsection{Electronic band structure}

\begin{figure}
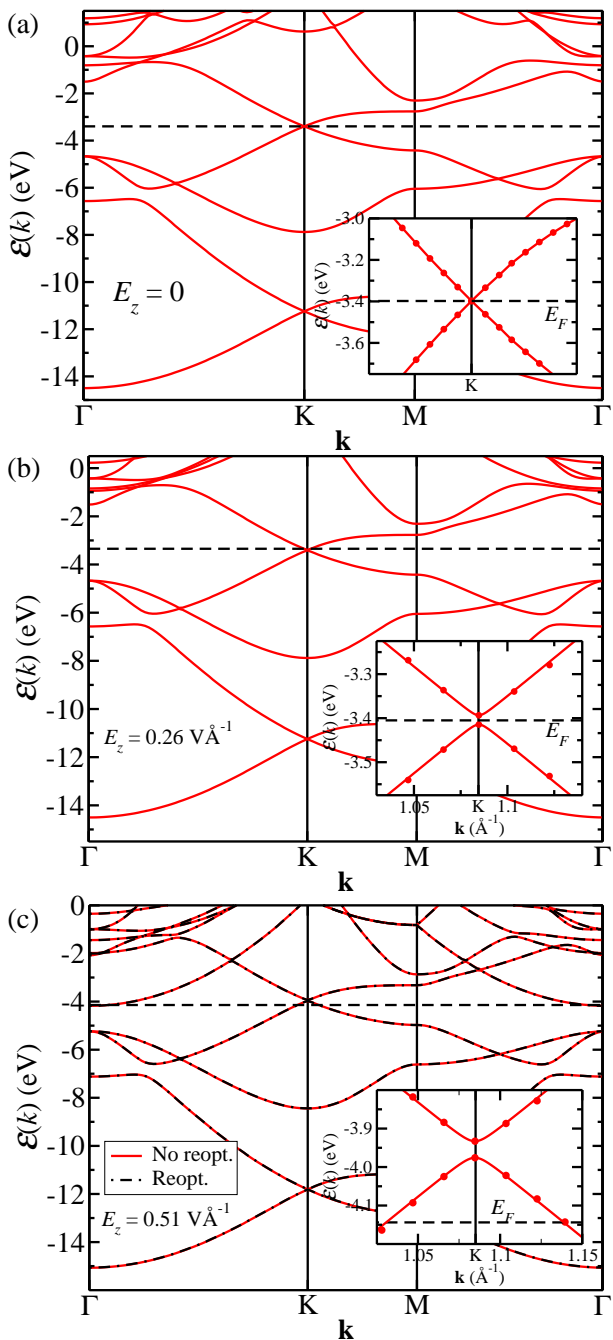

\begin{center}
\includegraphics[clip,width=0.45\textwidth]{mono_Ecut30_Nk53_Ez0_L50_PBE_BS.eps}
\\ \includegraphics[clip,width=0.45\textwidth]{mono_Ecut30_Nk53_Ez0.005_L50_PBE_BS.eps}
\\ \includegraphics[clip,width=0.45\textwidth]{mono_Ez0.01_Nk53_ecut30_L50.eps}
\caption{(Color online) DFT-PBE band structures for silicene in a cell of
  length $L_z=26.5$ {\AA} with a plane-wave cutoff energy of 816 eV and a
  $53\times 53$ ${\bf k}$-point grid: (a) in zero external electric field, (b)
  with $E_z=0.26$ V{\AA}$^{-1}$, and (c) with $E_z=0.51$ V{\AA}$^{-1}$ (shown
  both with and without the relaxation of the atomic coordinates in the
  electric field). The zero of the external potential is in the center of the
  silicene layer.  The dashed line shows the Fermi energy in each case and the
  insets show the spectrum near the Fermi level in the vicinity of the K
  point.
\label{fig:BS_v_Ez}}
\end{center}
\end{figure}

The calculated band structure of a ``free'' silicene layer is shown in
Fig.\ \ref{fig:BS_v_Ez}. As expected, it resembles the band structure of
graphene; in particular it shows the linear Dirac-type dispersion of electrons
near the K points, where we find the Fermi level in undoped silicene.  The
Fermi velocity $v$ of electrons in silicene is lower than that in graphene
(see Table \ref{table:parameters_summary}).  Although the lattice parameters
and sublattice buckling found in the different DFT calculations are in good
agreement, our results for the Fermi velocity are very much smaller than the
Fermi velocity reported in Ref.\ \onlinecite{cahangirov}.

\section{Application of a transverse electric
  field \label{sec:transverse_field}}

\subsection{Breaking the sublattice symmetry}

To exploit the weak buckling of silicene, we consider its behavior in an
external electric field $E_z$ applied in the $z$-direction, as shown in
Fig.\ \ref{fig:silicene_in_capacitor}.  The main effect of such an electric
field is to break the symmetry between the $A$ and $B$ sublattices of
silicene's honeycomb structure and hence to open a gap $\Delta$ in the band
structure at the hexagonal BZ points K and K$^\prime$. In the framework of a
simple nearest-neighbor tight-binding model, this manifests itself in the form
of an energy correction to the on-site energies that is positive for
sublattice $A$ and negative for $B$. This difference in on-site energies
$\Delta={\cal E}_A-{\cal E}_B$ leads to a spectrum with a gap for electrons in
the vicinity of the corners of the BZ: ${\cal E}_\pm=\pm
\sqrt{(\Delta/2)^2+|v{\bf p}|^2}$, where ${\bf p}$ is the electron ``valley''
momentum relative to the BZ corner.  Opening a gap in graphene by these means
would be impossible because the $A$ and $B$ sublattices lie in the same plane.

\subsection{First-order perturbation theory \label{sec:perturbation_theory}}

A na\"{\i}ve estimate of the electric-field-induced gap in silicene can be
made using first-order perturbation theory by diagonalizing a $2 \times 2$
Hamiltonian matrix at ${\bf p}\rightarrow{\bf 0}$,
\begin{equation} \delta {\cal H}(E_z) = eE_z \left[ \begin{array}{cc}
      \langle \psi_{\rm K}^- | z | \psi_{\rm K}^- \rangle & \langle \psi_{\rm
        K}^- | z | \psi_{\rm K}^+ \rangle \\ \langle \psi_{\rm K}^+ | z |
      \psi_{\rm K}^- \rangle & \langle \psi_{\rm K}^+ | z | \psi_{\rm K}^+
      \rangle \end{array} \right]. \end{equation} Here, $\psi_{\rm K}^\pm$ are
the degenerate lowest unoccupied and highest occupied Kohn-Sham orbitals at
the K point at $E_z=0$, and $z=0$ corresponds to the mid-plane of the buckled
lattice.  This suggests a band gap which opens linearly with the electric
field at a rate $d\Delta/dE_z=0.554$ and 0.573 e{\AA} for the wave functions
$\psi_K$ found using the LDA and PBE functionals, respectively.

\subsection{Self-consistent DFT calculations in the presence of the field}

\begin{figure}
\begin{center}
\includegraphics[clip,width=0.45\textwidth]{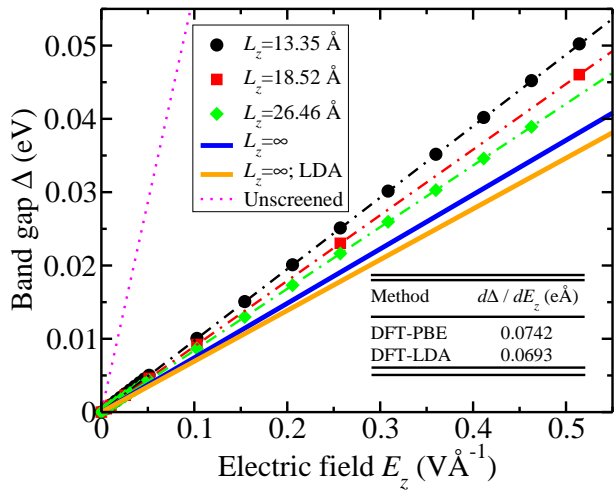}
\caption{(Color online) DFT gap against applied electric field $E_z$ for
  silicene with a plane-wave cutoff energy of 816 eV and a $53 \times 53$
  ${\bf k}$-point grid.  Unless otherwise stated, the PBE functional was
  used. The box length in the $z$ direction was varied from $L_z=13.35$ {\AA}
  to 26.46 {\AA}. The results have been extrapolated to the limit
  $L_z\rightarrow \infty$ of infinite box length (solid lines) as described in
  Sec.\ \ref{sec:boxlength}. Unscreened band gaps calculated using
  perturbation theory are also shown.  The inset table shows the calculated
  rate at which the band gap opens.
\label{fig:gap_v_Ez}}
\end{center}
\end{figure}

The estimate given in Sec.\ \ref{sec:perturbation_theory} is in fact only an
upper limit for the rate at which the band gap opens, since it neglects
screening by the polarization of the $A$ and $B$ sublattices. In order to
obtain an accurate value of the rate at which a band gap can be opened with an
electric field, we have performed fully self-consistent calculations of the
DFT band structure in the presence of an electric field. A typical result of
such a calculation is shown in Fig.\ \ref{fig:BS_v_Ez}(b).  At small electric
fields, relaxing the structure in the presence of the field does not have a
significant effect on the band gap, but the screening of the electric
potential by the sublattice polarization of the electron states makes a
substantial difference.  The DFT-calculated gaps are gathered in
Fig.\ \ref{fig:gap_v_Ez}.  The variation of the band gap $\Delta$ at K with
electric field $E_z$ is almost perfectly linear for fields up to $E_z \approx
1$ V{\AA}$^{-1}$. The results for the rate $d\Delta/dE_z$ at which a gap is
opened are shown in the table inset in Fig.\ \ref{fig:gap_v_Ez}. The eightfold
difference between the self-consistent and the unscreened values of
$d\Delta/dE_z$ indicates that the system exhibits a strong sublattice
polarizability.

Our value for the rate at which the band gap opens within DFT-PBE is 0.0742
e{\AA}. This is substantially lower than the result obtained by Ni\ \textit{et
  al.},\cite{ni2012} which is 0.157 e{\AA}.  Part of the reason for the
discrepancy is that we extrapolated our results to infinite box length,
whereas Ni\ \textit{et al.}\ used a fixed amount of vacuum between the
periodic images of the layers.  Another possible reason for the difference is
that we used a plane-wave basis set, whereas Ni \textit{et al.}\ used a
localized basis set.  An incomplete localized basis set would tend to
undermine the extent to which the electrons can adjust to screen the electric
field.

\subsection{Stability of the silicene lattice in an electric field}

The narrow-gap silicene band structure shown in Fig.\ \ref{fig:BS_v_Ez}
persists over a broad range of electric fields $E_z$.  However, for electric
fields of more than $E_z \approx 0.5$ V{\AA}$^{-1}$, the band gap starts to
close due to an overlap of the conduction band at $\Gamma$ and the valence
band at K, and silicene becomes a semimetal, as shown in
Fig.\ \ref{fig:BS_v_Ez}(c)\@.  According to our calculations, the buckled
honeycomb crystal is still metastable at this electric field, as can be seen
in Fig.\ \ref{fig:phonons_in_field}. The main effects of the electric field on
the phonon dispersion curve are (i) to lift some degeneracies at K and M and
(ii) to soften one of the acoustic branches, but without making the frequency
imaginary. Under much higher electric fields, the honeycomb structure of
silicene becomes unstable. We found that $E_z \geq 2.6$ V{\AA}$^{-1}$ causes
the lattice parameter to increase without bound when the structure is relaxed.

\section{SO coupling in silicene \label{sec:spin_orbit}}

\subsection{SO-induced gap}

\begin{figure}
\begin{center}
\includegraphics[clip,width=0.45\textwidth]{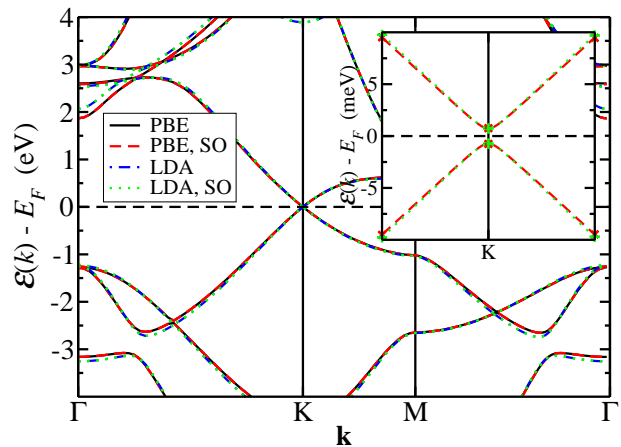}
\caption{(Color online) DFT-PBE and DFT-LDA band structures with and without
  SO coupling taken into account. The inset shows the bands around the
  K-point, revealing a small band gap induced by SO coupling. The width of the
  bottom panel corresponds to $1/200$ of the $\Gamma$K
  line. \label{fig:SO_gap}}
\end{center}
\end{figure}

We have also performed a study of the effects of SO coupling (which is more
pronounced in Si than in C) on the band structure. The SO coupling term is
explicitly included in the Hamiltonian in the DFT calculations.  The results
obtained with the LDA and PBE functionals are shown in
Fig.\ \ref{fig:SO_gap}. Both functionals predict an SO gap of the order of a
few meV at the K point, while the rest of the band structure barely differs
from the nonrelativistic case.  Our calculated LDA and PBE SO gaps are 1.4 meV
and 1.5 meV, respectively, in agreement with the recent
literature.\cite{liu_2011}

\subsection{Crossover from topological to band insulating behavior}

In the theory of Dirac electrons on the honeycomb lattice, the SO gap is
accounted for by the Kane-Mele term describing, e.g., intrinsic SO coupling in
graphene.\cite{kane_mele}  The Kane-Mele SO coupling and the electric-field
induced $A$-$B$ sublattice asymmetry for electrons in the vicinity of the BZ
corners ${\rm K}_\pm=(\pm 4 \pi/(3a),0)$ in silicene can be incorporated in
the Hamiltonian
\begin{equation} H_{{\rm K}_\pm} = v {\bf p} \cdot {\bm \sigma}+ \Delta_{\rm SO}
  s_z \sigma_z+\frac{1}{2} \xi \Delta_z
  \sigma_z, \label{eqn:so_hamiltonian} \end{equation} where $\xi=\pm 1$
distinguishes between the two valleys, ${\rm K}_+$ and ${\rm K}_-$, in
silicene's spectrum.  Here, the Pauli matrices $\sigma_x$, $\sigma_y$, and
$\sigma_z$ act in the space of the electrons' amplitudes on orbitals
attributed to the $A$ and $B$ sublattices, $(\psi_A,\psi_B)$ for the valley at
${\rm K}_+$ and $(\psi_B,-\psi_A)$ for the valley at ${\rm K}_-$.  In
Eq.\ (\ref{eqn:so_hamiltonian}), $s_z$ is the electron spin operator normal to
the silicene plane, and $\Delta_{\rm SO}$ and $\Delta_z$ are the
DFT-calculated SO-coupling and electric-field induced gaps.

The Hamiltonian of Eq.\ (\ref{eqn:so_hamiltonian}) generically describes the
transition between the 2D topological and band-gap insulators.  Its spectrum,
\begin{eqnarray} {\cal E}_{\uparrow \pm} & = & \pm \sqrt{ \frac{1}{4} \left(
    \Delta_{\rm SO}+\xi \Delta_z \right)^2+v^2p^2}, \nonumber \\ {\cal
    E}_{\downarrow \pm} & = & \pm \sqrt{ \frac{1}{4} \left( \Delta_{\rm
      SO}-\xi \Delta_z \right)^2+v^2p^2}, \end{eqnarray} includes two gapped
branches, one with a larger gap $|\Delta_{\rm SO}+\Delta_z|$ and another with
a smaller gap $|\Delta_{\rm SO}-\Delta_z|$.  At a critical external electric
field $E_z^c \approx 20$ mV{\AA}$^{-1}$, $\Delta_{\rm SO}=\Delta_z$, and the
smaller gap closes, marking a transition from a topological
insulator\cite{hasan_kane,qi_2011,kane_mele} at $\Delta_{\rm SO}>\Delta_z$ to
a simple band insulator at $\Delta_{\rm SO}<\Delta_z$.  The difference between
these two states of silicene is that the topological insulator state supports
a gapless spectrum of edge states for the electrons, in contrast to a simple
insulator, where the existence of gapless edge states is not protected by
topology.  However, one may expect something reminiscent of the topological
properties of Dirac electrons to show up even in the band insulator state of
silicene: an interface between two differently gated regions, with electric
fields $E_z$ and $-E_z$ (where $E_z \gg E_z^c$), should support a
one-dimensional gapless band with an almost linear dispersion of
electrons.\cite{semenoff}

\section{Computational details \label{sec:comp_meth}}

\subsection{Cohesive energy}

All our plane-wave DFT total energies were corrected for finite-basis
error\cite{francis} and it was verified that the residual dependence of the
total energy on the plane-wave cutoff energy is negligible. We used ultrasoft
pseudopotentials throughout, except where otherwise stated.  The silicene
system was made artificially periodic in the $z$-direction (normal to the
silicene layer) in our calculations.  The atomic structure was obtained by
relaxing the lattice parameter and atom positions within DFT, subject to the
symmetry constraints and at fixed box length $L_z$ in the $z$ direction. The
cohesive energy was then evaluated using this optimized structure.

The energy of an isolated Si atom (needed when evaluating the cohesive energy)
was obtained in a cubic box of side-length $L$ subject to periodic boundary
conditions.  We extrapolated the energy of the isolated atom to the limit of
infinite box size by fitting
\begin{equation} E(L)=E(\infty)+cL^{-8} \label{eq:boxlength} \end{equation}
to the DFT energies $E(L)$ obtained in a range of box sizes, where $E(\infty)$
and $c$ were parameters determined by fitting.  Equation (\ref{eq:boxlength})
gave a very good fit to our data.

We have also calculated the DFT zero-point correction to the energy of
silicene. This is expected to be largely independent of the
exchange-correlation functional used. Indeed, our calculations show that the
zero-point correction is 0.103 eV within the LDA and 0.101 eV with the PBE
functional.\cite{pbe}  We used the PBE result in our final calculations of the
cohesive energy reported in Table \ref{table:parameters_summary}.

\subsection{Evaluation of the Fermi velocity}

To evaluate the Fermi velocity shown in Table \ref{table:parameters_summary}
we evaluated the DFT band structure using a $53\times 53$ ${\bf k}$-point grid
and a plane-wave cutoff energy of 816 eV in a cell of length $L_z=26.46$
{\AA}.  We then fitted Eq.\ (17) of Ref.\ \onlinecite{winkler} to the highest
occupied and lowest unoccupied bands within a circular region around the K
point; the Fermi velocity is one of the fitting parameters.  The radius of the
circular region was 6\% of the length of the reciprocal lattice vectors; we
verified that the Fermi velocity was converged with respect to this radius.

\subsection{Geometry optimization and phonon dispersion curves}

The phonon dispersion curves shown in Sec.\ \ref{sec:phonon_results} were
calculated using the method of finite displacements, with atom displacements
of 0.042 {\AA}, in a supercell consisting of $3\times 3$ primitive cells with
a $20 \times 20$ ${\bf k}$-point grid in the primitive cell.  In the results
with the external electric field, the box length was $L_z=19.05$~{\AA} and the
plane-wave cutoff energy was 435 eV\@. In the results without the field, the
box length was $L_z=13.35$ {\AA} and the plane-wave cutoff was 816 eV\@.  This
choice was made because the error due to a finite box length $L_z$ is
potentially much larger in the presence of a transverse electric field.

\begin{figure}
\begin{center}
\includegraphics[clip,width=0.45\textwidth]{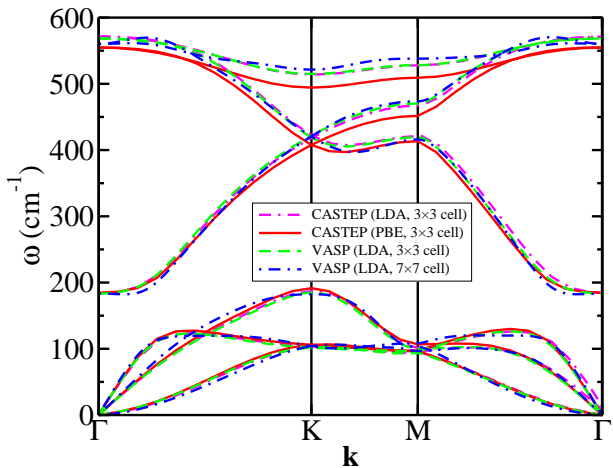}
\caption{(Color online) Phonon dispersion curves for silicene obtained with
  \textsc{castep} and \textsc{vasp} using different exchange-correlation
  functionals and supercell sizes.  The results for a $3 \times 3$ supercell
  were obtained with a box length $L_z=13.35$ {\AA}, a $20\times 20$ ${\bf
    k}$-point grid in the primitive cell, and a plane-wave cutoff energy of
  816 eV\@. The matrix of force constants was evaluated using the method of
  finite displacements, with the displacements being 0.042 {\AA}.  The results
  for a $7 \times 7$ supercell were obtained with a box length $L_z=15.0$
  {\AA}, a $12\times 12$ ${\bf k}$-point grid in the supercell, and a
  plane-wave cutoff energy of 500 eV\@.  The matrix of force constants was
  evaluated using the method of finite displacements, with displacements of
  0.09 {\AA}. \label{fig:mono_phonons}}
\end{center}
\end{figure}

The geometry optimization and band-structure calculations at zero external
field were performed with both the \textsc{castep}\cite{castep,castep_dfpt}
and \textsc{vasp}\cite{vasp} codes, to verify that the results are in good
agreement. This check was necessary because it was only possible to perform
the electric-field calculations with \textsc{castep}, while for the SO
calculations we had to use \textsc{vasp}. In principle the only difference
between the calculations performed using the two codes arises from the Si
pseudopotentials used.  The PAW method\cite{blochl} was used in the
\textsc{vasp} calculations, whereas ultrasoft pseudopotentials were used in
the \textsc{castep} calculations.  As can be seen in Table
\ref{table:parameters_summary}, the geometries predicted by the two codes
agree well. We have also verified that the band structures are in good
agreement.  Finally, in Fig.\ \ref{fig:mono_phonons} we show that the phonon
dispersions obtained with the two codes are virtually identical when the same
parameters are used.

Figure \ref{fig:mono_phonons} also demonstrates that our phonon dispersion
curves are converged with respect to supercell size.

\subsection{Band gap in the presence of an external
  electric field}

\subsubsection{Plane-wave cutoff energy}

\begin{figure}
\begin{center}
\includegraphics[clip,width=0.45\textwidth]{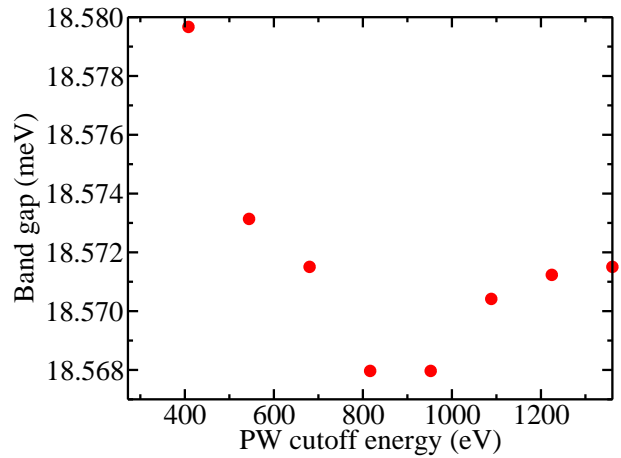}
\caption{(Color online) DFT-PBE gap against plane-wave (PW) cutoff energy for
  silicene subject to an electric field of 0.257 V/{\AA} in a cell of length
  $L_z=13.35$ {\AA} with a $15\times 15$ ${\bf k}$-point grid including
  K\@. \label{fig:gap_v_ecut}}
\end{center}
\end{figure}

The convergence of the calculated band gap with respect to the plane-wave
cutoff energy for a particular applied field is shown in
Fig.\ \ref{fig:gap_v_ecut}.  The gap converges extremely rapidly.

\subsubsection{{\bf k}-point sampling \label{sec:kpoints}}

\begin{figure}
\begin{center}
\includegraphics[clip,width=0.45\textwidth]{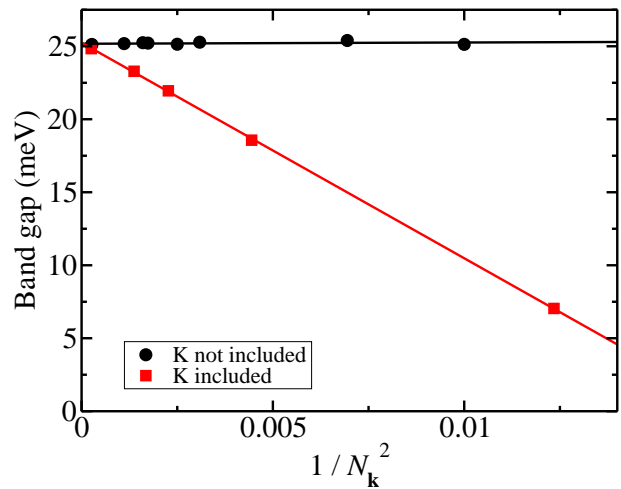}
\caption{(Color online) DFT-PBE gap against the reciprocal of the number
  $N_{\bf k}^2$ of ${\bf k }$-points for silicene subject to an electric field
  of 0.257 V/{\AA} in a cell of length $L_z=13.35$ {\AA} with a plane-wave
  cutoff energy of 816 eV\@.  \label{fig:gap_v_Nk}}
\end{center}
\end{figure}

The convergence of the calculated band gap at BZ point K with respect to the
${\bf k}$-point grid used in the self-consistent field calculations is shown
in Fig.\ \ref{fig:gap_v_Nk}.  The finite-sampling error falls off as the
reciprocal of the total number of ${\bf k}$ points.  The prefactor of the
finite-sampling error is vastly greater when K or K$^\prime$ is included in
the grid of ${\bf k}$-points for the self-consistent field calculations.

\subsubsection{Choice of pseudopotential}

\begin{figure}
\begin{center}
\includegraphics[clip,width=0.45\textwidth]{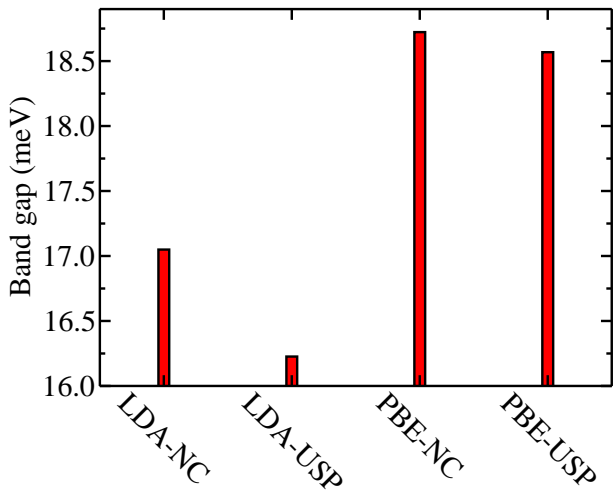}
\caption{(Color online) DFT gap with different exchange-correlation
  functionals (LDA and PBE) and pseudopotentials [on-the-fly ultrasoft (USP)
    and norm conserving (NC)\cite{castep}] for silicene subject to an electric
  field of 0.257 V/{\AA} in a cell of length $L_z=13.35$ {\AA} with a
  $15\times 15$ ${\bf k}$-point grid and a plane-wave cutoff energy of 816
  eV\@. \label{fig:gap_v_psp}}
\end{center}
\end{figure}

The dependence of the calculated gap on the exchange-correlation functional
and pseudopotential is shown in Fig.\ \ref{fig:gap_v_psp}.  The difference
between the results obtained with different pseudopotentials is much smaller
than the gap, but is not wholly negligible. The on-the-fly ultrasoft
pseudopotential is believed to be more accurate than the norm-conserving
pseudopotential,\cite{castep} and hence we have used the former in our final
calculations.

\subsubsection{Box length \label{sec:boxlength}}

\begin{figure}
\begin{center}
\includegraphics[clip,width=0.45\textwidth]{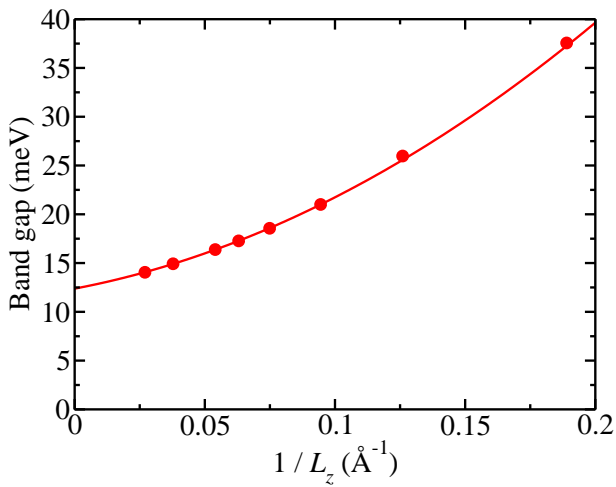}
\caption{(Color online) DFT-PBE gap against the reciprocal of the box-length
  $L_z$ for silicene subject to an electric field of 0.257 V/{\AA} with a
  $15\times 15$ ${\bf k}$-point grid and a plane-wave cutoff energy of 816
  eV\@.  The solid line is a fitted quadratic in $1/L_z$. \label{fig:gap_v_L}}
\end{center}
\end{figure}

The dependence of the calculated gap on the length of the simulation box is
shown in Fig.\ \ref{fig:gap_v_L}.  It is clear that this is potentially a
large source of error.  However, the $L_z$-dependence is reasonably well
approximated by a quadratic in $1/L_z$, allowing the DFT gaps to be
extrapolated to infinite cell size if results at three or more different cell
sizes are available.

\subsubsection{Estimates of uncertainty in our final results}

Our final results for the field-induced gap were obtained using $53 \times 53$
${\bf k}$ points, not including K or K$^\prime$, a plane-wave cutoff energy of
$E_{\rm cut}=816$ eV, and box lengths $L_z=13.35$, 18.521, and 26.459 {\AA};
the results were then extrapolated to infinite box length by fitting a
quadratic in $1/L_z$.  From the magnitudes of the variations shown in
Figs.\ \ref{fig:gap_v_ecut}--\ref{fig:gap_v_L}, we estimate the uncertainty in
our final results for the rate $d\Delta/dE_z$ at which the band gap opens when
an electric field is applied to be less than about 0.01 e{\AA}\@.

\subsection{Unscreened estimate of the band gap in
  the presence of an external electric field}

To evaluate the field-induced band gap using perturbation theory we used
norm-conserving pseudopotentials.\cite{castep} We used a $39 \times 39$ ${\bf
  k}$-point mesh including the K point, and a cell length of $L_z=26.46$
{\AA}.  It was verified that the perturbation-theory-induced rate of gap
opening $d\Delta/dE_z$ was converged to within 0.00002 e{\AA} with respect to
${\bf k}$-point mesh and $L_z$.  The finite-basis error in $d\Delta/dE_z$ was
found to fall off approximately exponentially with respect to the plane-wave
cutoff energy, and hence we extrapolated our results to basis-set
completeness.

\subsection{Band structure with SO coupling}

The SO calculations were performed with a plane-wave cutoff of 500 eV and a
$24 \times 24$ ${\bf k}$-point grid.  We checked that the length of the
simulation box has negligible influence on the SO gap: the gap is the same
with simulation box lengths of 15 {\AA} and 30 {\AA} up to numerical accuracy.

\section{Conclusions \label{sec:conclusions}}

In summary, we have shown that a 2D layer of Si atoms---silicene---is a
versatile material in which a band gap can be tuned (in a broad range of tens
of meV) using a transverse electric field $E_z$, while silicene remains
metastable.  At the low field $E_z \approx 20$ mV{\AA}$^{-1}$, we expect
silicene to undergo a transition between a topological and a simple band
insulator, whereas at much higher field $E_z\approx 0.5$ V{\AA}$^{-1}$ it will
undergo a transition from a band insulator into a semimetal.

\begin{acknowledgments}
We acknowledge financial support from the EPSRC through a Science and
Innovation Award, the EU through the grants Concept Graphene and CARBOTRON,
the Royal Society, and Lancaster University through the Early Career Small
Grant Scheme.  Computational resources were provided by Lancaster University's
High-End Computing facility. We thank H.-J.\ Gao and G.\ Le Lay for useful
discussions, and J.\ R.\ Wallbank for providing us with
Fig.\ \ref{fig:silicene_in_capacitor}.
\end{acknowledgments}

\end{document}